\documentclass[12pt]{article}
\usepackage[polish,english]{babel}
\usepackage[latin1]{inputenc}
\usepackage{times}
\usepackage[T1]{fontenc}
\usepackage{epsfig}

\begin{document}

\title{Compact Q-balls and Q-shells in a scalar electrodynamics }

\author{H. Arod\'z $\;$ and $\;$ J. Lis\\$\;\;$ \\  Institute of Physics,
Jagiellonian University, \\ Reymonta 4, 30-059 Cracow, Poland}

\date{$\;$}

\maketitle

\begin{abstract}
We investigate spherically symmetric non topological solitons in electrodynamics with a  scalar field self
interaction $U \sim |\psi|$ taken from the complex signum-Gordon model. We find Q-balls for small absolute
values of the total electric charge $Q$, and Q-shells when $|Q|$ is large enough. In both cases the charge
density exactly vanishes outside certain compact region in the three dimensional space. The dependence of the
total energy $E$ of small Q-balls  on the total electric charge has the form $E \sim |Q|^{5/6}$, while in the
case of very large Q-shells $E \sim |Q|^{7/6}$.
\end{abstract}

\vspace*{2cm} \noindent PACS: 11.27.+d, 98.80.Cq, 11.10.Lm \\

\pagebreak

\section{ Introduction}

Several field-theoretic models predict existence of stationary
finite energy excitations called Q-balls, see \cite{1, 2} for a
review. Such objects arise as a highly nontrivial consequence of
collective nonlinear dynamics of the fields.  In contradistinction
to topological defects which appear because of a frustration of the
field interpolating between different degenerate vacua, the Q-balls
exist in models with non degenerate vacuum. They must be taken into
account in every thorough analysis of the dynamics of the fields.

In a recent paper \cite{3} we have shown that the Q-balls can be
found also in the complex signum-Gordon model.   They differ from
other Q-balls by their strictly finite size: the vacuum field $\psi
=0$ outside a spherically symmetric Q-ball is reached at a finite
radius $R$, and it is approached in a parabolic manner. Such
parabolic behavior is the general feature of field theoretic models
with V-shaped self interaction terms \cite{4}. It is a consequence
of the fact that the field theoretic force density, which in the
case of the complex signum-Gordon model with the complex scalar
field $\psi$ is given by $ -
\partial U/\partial \psi^*$,  where $U =\lambda|\psi|$ and $\lambda
>0$ is the self coupling constant,  does not vanish in the limit $\psi
\rightarrow 0$. Therefore, for weak fields it is always stronger than the gradient force ($\sim \triangle \psi$)
which alone would lead to an infinitely long tail in the spatial asymptotics of the field, and also stronger
that the mass term ($\sim m^2 \psi$) which together with the gradient force would lead to the well-known
exponential asymptotics for the field. Heuristically, one can say that the parabolic behavior is due to the fact
that the field resists with a finite force density even the slightest deviations from the vacuum configuration.
The  self interaction term $U(|\psi|) = \lambda |\psi|$  is V-shaped because its plot has the form of the
inverted symmetric cone with the vacuum field $\psi=0$ right at the tip. The models with V-shaped potentials are
interesting for several reasons \cite{4}. Let us mention here a scaling symmetry of the on-shell type, and the
lack of a free field regime because the field equations can not be linearized around the vacuum solution. In
case of smooth self interactions the second derivative of pertinent field potential at $\psi =0$ gives a mass
scale  which is physically important in the vacuum sector. Defined in this manner mass scale is infinite if the
potential is V-shaped.

The motivation for the present paper comes from the idea that the
finite vacuum restoring force can perhaps win over the electrostatic
repulsion for arbitrarily large electrically charged Q-balls. In
order to address this question we extend the complex signum-Gordon
model by including an Abelian gauge field $A_{\mu}$ minimally
coupled to the scalar field. Thus, the global $U(1)$ symmetry is
replaced by the local one. The considered model is a version of
scalar electrodynamics.

It has turned out that  in such  scalar electrodynamics there exist electrically charged Q-balls provided that
the absolute value of the total electric charge $Q$ is small enough. For larger values of $|Q|$ we find, rather
surprisingly, Q-shells with completely empty interior. Ther is no upper bound on $|Q|$.   Both Q-balls and
Q-shells are static, spherically symmetric, and the charge density exactly vanishes outside certain finite,
$Q$-dependent radius $R$. The scalar field approaches its vacuum value $\psi=0$ in the parabolic manner, while
the electric field  of course has the standard Coulomb tail. The total energy is finite. Another very
interesting finding is a zigzag in the plot of the total energy versus the total electric charge $Q$ at
intermediate values of $Q$, where the Q-balls transform into the Q-shells.

In literature one can find several other versions of scalar electrodynamics with Q-balls, see \cite{5, 6, 7, 8}.
In all of  them there exists a maximal value of the total electric charge the static Q-balls can have.  Q-shells
were not found. In our case the electric charge can be arbitrarily large, but there is the transition from
Q-balls to Q-shells at certain intermediate value of the charge.

The plan of our paper is as follows. In Section 2 we introduce the model and  discuss  field equations for the
spherical solitons. The Q-ball solutions for small values of $|Q|$ are presented in Section 3.  Section 4 is
devoted to the Q-shell solutions in the case of large values of $|Q|$. The transition from Q-balls to Q-shells
at the intermediate values of $|Q|$ is studied in Section 5.  Section 6 contains a summary and remarks.

\section{Preliminaries }

The Lagrangian of our model has the  following form
\begin{equation}
L = - \frac{1}{4} F_{\mu\nu} F^{\mu\nu} + (D_{\mu}\psi)^* D^{\mu} \psi - \lambda |\psi|,
\end{equation}
where $\psi$ is the complex scalar field in $(3+1)$-dimensional Minkowski space-time, $\lambda >0$ is the self
coupling constant, $|\psi|$ is the modulus of $\psi$, $D_{\mu} \psi = \partial_{\mu} \psi + i q A_{\mu} \psi$,
where $q>0$ is the electromagnetic coupling constant and $A_{\mu}$ is the $U(1)$ gauge field.   For the sake of
convenience the fields $\psi, \; A_{\mu}$, the space-time coordinates $x^{\mu}$ and the constants $\lambda, \;q$
are dimensionless. Of course, in physical applications they should be multiplied by certain dimensional
constants. Euler-Lagrange equations give the Maxwell equations
\begin{equation}
\partial_{\mu}F^{\mu\nu} = j^{\nu}
\end{equation}
with the current density
\begin{equation}
 j^{\nu} = i q \left(\psi^* \partial^{\nu} \psi - \partial^{\nu} \psi^* \psi \right) - 2 q^2 A^{\nu} \psi^* \psi,
\end{equation}
and  the scalar field equation
\begin{equation}
D^{\mu} D_{\mu} \psi= - \frac{\lambda}{2} \frac{\psi}{|\psi|}.
\end{equation}
By definition, the r.h.s. of Eq. (4) is equal to 0 when $\psi =0$ (see Section II in \cite{3}). The field
potential $U(|\psi|) = \lambda |\psi|$  can be regarded as the limit of the smooth potential $U_a(|\psi|) =
\lambda \sqrt{a+|\psi|^2}$ when $a \rightarrow 0+$.

We will consider the simplest static, spherically symmetric Q-balls with vanishing magnetic field. The
corresponding Ansatz has the form
\begin{equation}
\psi = e^{i \omega t} F(r), \;\; \vec{A} =0, \;\; A_0 = A_0(r).
\end{equation}
Here $F(r)$ is a real function of the radial coordinate $r$ and $\omega >0$ is a constant real frequency. The
spatial part of the Maxwell equations is trivially satisfied, while the Gauss law has the form
\begin{equation}
\triangle A_0 = 2 q (\omega + q A_0) F^2,
\end{equation}
where $\triangle$ is the three dimensional Laplacian. The scalar
field equation is reduced to
\begin{equation}
\triangle F + (\omega + q A_0)^2 F = \frac{\lambda}{2} \mbox{sign} F,
\end{equation}
where $\mbox{sign}(0) =0$. Equations (6), (7) acquire a simpler form when we change a little bit the notation.
With
\[
\kappa = \frac{\lambda q}{\sqrt{2}}, \;\; G = \sqrt{2} q F, \;\; B = \omega + q A^0, \] we obtain the following
equations
\[
\triangle B = B G^2, \;\;\; \triangle G = - G B^2  + \kappa \:
\mbox{sign} G.
\]
Note that the fields  $B,\: G$ are gauge-invariant as opposed to $A_0, \:\psi$ and $\omega$ (the Ansatz (5)
fixes the gauge only partially because the $U(1)$ gauge transformations which depend only on time are still
allowed).

 Furthermore, the parameter $\kappa$ can be removed with
the help of  rescaling
\[
B(r) \rightarrow \kappa^{1/3} B(\kappa^{1/3} r), \;\;\; G(r) \rightarrow \kappa^{1/3} G(\kappa^{1/3} r),  \;\;\;
r \rightarrow \kappa^{1/3} r,
\]
and therefore we may put $\kappa =1$ without any loss of generality. After taking into account the spherical
symmetry we finally obtain the following equations
\begin{equation}
B'' = - \frac{2}{r} B' + B G^2,
\end{equation}
\begin{equation}
G'' = - \frac{2}{r} G' - G B^2 + \mbox{sign}G,
\end{equation}
where $'$ denotes the derivative $d/dr$.  These equations are supplemented by two conditions. First, continuity
of $\nabla \psi, \; \nabla A_0$ at $r=0$ implies the conditions
\begin{equation}
B'(0) = 0 = G'(0).
\end{equation}
Second, because Eqs. (8), (9) together with the condition (10) are
invariant under the transformations
\[
B(r) \rightarrow - B(r), \;\;\; G(r) \rightarrow - G(r),
\]
we may also assume that
\begin{equation}
B(0) \geq 0, \;\; G(0) \geq 0.
\end{equation}
We shall see that in fact $B(0)>0$.

The total electric charge $Q$ and the total energy $E$ are given by the following formulas
\begin{equation}
Q = \int d^3x \; j^0 = - \frac{4\pi}{q} \int_0^{\infty}\!\! dr \;
r^2 B G^2,
\end{equation}
\[E = \int d^3x \; \left[ \frac{1}{2} (\nabla A_0)^2 +
|D_0\psi|^2 + \nabla\psi^* \nabla\psi + \lambda |\psi| \right] = \frac{2\pi}{q^2} \underline{E}, \] where
\begin{equation} \underline{E}  =  \int_0^{\infty}\!\! dr \; r^2\: [B^{'2} + G^{'2} +B^2 G^2 + 2
\kappa |G| ].
\end{equation}
As already signalled, we put $\kappa =1$.

The Q-balls and Q-shells are represented by solutions of Eqs. (8), (9) which obey the conditions (10), (11) and
have finite $Q$ and $E$. Let us have a look at Eq. (8). It can be written in the integral form
\[
B'(r) = \frac{1}{r^2} \int_0^r\!\! dr'\: r^{'2} B(r') G^2(r'),
\]
which implies that the function $B(r)$ can not decrease when $r$
increases because $ B'(0)\geq 0$. Furthermore, we shall shortly see
that in the case of Q-balls and Q-shells $G(r)$ vanishes for all $ r
> R$, where $R$ is the finite radius of the cloud of the electric
charge. Therefore, $B'(r) = \underline{Q} / r^2$ for $r >R$, where \begin{equation} \underline{Q} = \int_0^R dr
\: r^2 B G^2 = - \frac{q Q}{4 \pi},
\end{equation}
and in consequence, $B(r) = \beta - \underline{Q} /r$ for large $r$. The usual condition that the electrostatic
potential $A_0$ vanishes at the spatial infinity and  the definition of the function $B$ imply that $\beta =
\omega$. Hence, for $r
>R$
\begin{equation}
B(r) = \omega - \frac{\underline{Q}}{r}.
\end{equation}
It follows that $A_0(r) = Q/(4 \pi r)$ for large $r$, as expected for the static spherically symmetric,
spatially localized electric charge distribution. \footnote{Note that with our sign conventions the total
electric charge $Q $ is negative while $\underline{Q} >0$. We call $\underline{Q}$ the charge and $Q$ the total
electric charge.}

It is the well-known fact that   Q-ball solutions in models with a global $U(1)$ symmetry minimize the total
energy $E$ under the condition that the U(1) global charge is kept constant \cite{1, 9}. Also in the considered
case of local $U(1)$ gauge symmetry our Q-balls and Q-shells minimize the total energy, but one has to assume
not only that the total electric charge $Q$ is constant but also that the Gauss law (8) holds. This latter
condition follows from the fact that the Gauss law is in fact a constraint which is essential for the physical
content of the electrodynamics. Therefore the variations $\delta B$, $\delta G $ have to be interrelated in such
a manner that the fields $B+\delta B$, $G+\delta G$ also obey that constraint. It follows that in the linear
approximation the admissible spherically symmetric variations obey the condition
\[
\delta B '' = - \frac{2}{r} \delta B ' + G ^2 \delta B + 2 B G \:
\delta G. \] Using this constraint one can show that
\[
\delta \underline{E} = 2 \omega \: \delta \underline{Q} - 2
\int_0^{\infty} \!\! dr r^2\: \left[G'' + 2 G'/r + B^2 G -
\mbox{sign}(G)\right]\; \delta G. \] It is clear that the conditions
$\delta \underline{E}=0, \; \delta \underline{Q}=0$ imply Eq. (9).
Moreover, it turns out that the second variation of the energy is
positive because it can be written in the form
\[ \delta^2 \underline{E} = \int_0^{\infty} \!\! dr\: r^2\: \left[
(\delta B')^2 + (\delta G')^2 + (G \delta B + B \delta G)^2 +
\frac{a (\delta G)^2}{(a+ G^2)^{3/2}} \right] \] (here we use the
regularized version $U_a(|\psi|)$ of the field potential). Thus, the
Q-ball and Q-shell solutions indeed give local minima of the energy.

\section{The Q-balls}

Qualitative analysis of  solutions of the set  of Eqs. (8), (9) can be accomplished  with the help of the
standard tool: a mechanical analogy.  In the context of Q-balls such interpretation of field equations was used
already in \cite{9}. Thus, we regard (8), (9) as Newton's equations of motion for a fictitious particle of unit
mass moving in the $(B, G)$-plane. Because of the assumptions (11)  we actually consider  motions in the first
quarter of the plane. The radial coordinate $r$ plays the role of time. There are two forces acting on the
particle: the friction force given by the two-vector $( - 2 B'/r, -2 G'/r)$ and the external force $ \vec{f} =
(BG^2, \mbox{sign} G - B^2 G)$. The external force does not have a potential, i.e., $\vec{f} \neq - \nabla u$
with certain $u$. This fact reflects the different origins of Eq. (8) and (9): Eq. (8) is in fact the Gauss law
for the electric field, hence it is a constraint, while Eq. (9) comes from the dynamical evolution equation for
the scalar field. The $B$ component of $\vec{f}$ (equal to $B G^2$) is nonnegative. This means that the particle
is constantly pushed in the $B$ direction until $G=0$. The $G$ component of $\vec{f}$ vanishes along the
hyperbola $G B^2 =1$ and along the straight line $G=0$. It is negative above that hyperbola (a `northern slope')
and positive when $ 0 < G < 1/B^2$ (a `southern slope'). One may imagine that the particle moves in a curved
valley with the bottom on the hyperbola, and that there is a `$B$-wind' which constantly pushes the particle in
the positive $B$ direction.

The mechanical analogy presented above might seem relatively complicated, nevertheless it gives the right
intuitions about the existence and the form of solutions of Eqs. (8), (9). For example, one can guess that
perhaps the `B-wind' can just push the particle along the bottom of the valley. This would mean that there
exists a solution such that $G(r) B^2(r) =1$. Indeed, it turns out that this is the case: the solution has the
form
\[
B(r) = \beta_0 \sqrt{r}, \;\; G(r) = \frac{\gamma_0}{r},
\]
where $\beta_0^2 = 2/\sqrt{3}, \; \gamma_0 = \sqrt{3}/2$.
Unfortunately, this solution is not interesting as far as the
Q-balls are concerned because it is singular at $r=0$ and it has the
infinite total energy and charge.

Note that there are no Q-balls such that $B(0) =0$.  In this case Eq. (8) and the condition $B'(0)=0$  imply
that $B(r) =0$ for all $r \geq 0$, i.e., that the particle moves along the $G$-axis. Then, it follows from Eq.
(9) that $G(r) = G(0) + r^2/6$. The electric charge density for this solution vanishes, but the total energy is
infinite.

The Q-ball solution corresponds to the trajectory of the particle which starts at the `time' $r=0$ with the
vanishing initial velocity $B'(0)= G'(0)=0$  from a point $(B(0), G(0))$ lying on the `northern slope', and
reaches the $B$-axis at a finite `time' $R$, $G(R)=0$. If at that `time' also $G'(R)=0$, then for $r >R$ the
particle moves along the $B$-axis, asymptotically approaching the point $(B, G) = (\omega, 0)$, as it follows
from formula (15). It is stopped by the friction force, while the external force $\vec{f}$ vanishes on the
$B$-axis. The condition $G'(R)=0$ means that the particle lands on the $B$-axis tangentially to it. This is
possible because of the `$B$-wind'. The trajectories of the type  described above correspond to the simplest
Q-balls. More complicated solutions can exist too, see the remark 4 in Section 6.

We have found the solutions of Eqs. (8), (9) corresponding to  the above discussed  trajectories numerically.
For a given value of $B(0) >0$ we have adjusted the value of $G(0) > 1/B^2(0)$ until $G(R)=0=G'(R)$ for a
certain value of $R$. Example of such a numerical solution is shown in Fig. 1.
\begin{center}
\begin{figure}[tph!]
\hspace*{1cm}
\includegraphics[height=7cm, width=11cm]{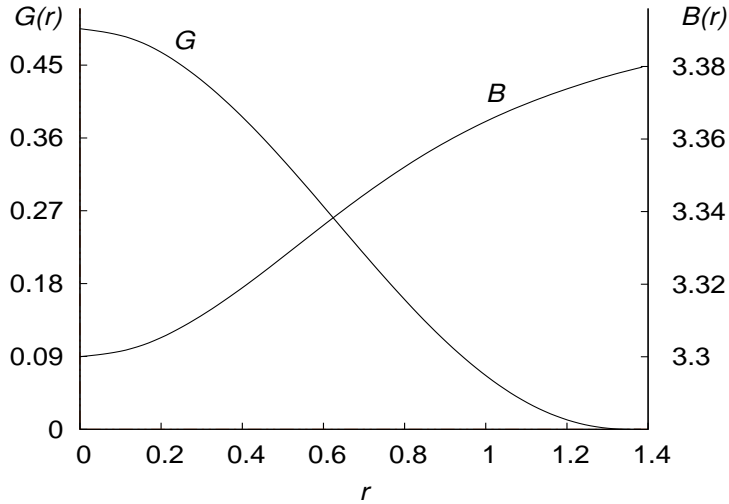}
\caption{Example of the Q-ball solution.  For this solution $B(0)$ =3.30, $\:G(0)=$ 0.49...,  $R= $1.35...,
$B(R)=$3.37..., $\underline{Q} =$ 0.05..., $\underline{E}= $0.44..., $\omega =$ 3.41... .  }
\end{figure}
\end{center}

The numerical analysis is nicely supplemented by an analytical
approximation which works in the case of very large values of
$B(0)$. It is based on the observation that in this case the
`northern slope' is very steep and that the bottom of the  valley
lies very close to the $B$-axis. Therefore, the particle very
quickly reaches the bottom of the valley and climbs the `southern
slope'. The  `$B$-wind' is relatively weak on that slope because it
is proportional to $G^2$, see the r.h.s. of Eq. (8). Thus, we may
expect that the coordinate $B$ of the particle is almost constant.

Let us assume that $B(r)$ is just constant, $B(r) = B(0)$ for $0 \leq r \leq R$ and that $B(0)$ is very large.
Then Eq. (8) is approximately satisfied: because $ GB^2\approx 1$ close to the bottom of the valley, the term
$BG^2$ is of the order $B^{-3}(0)$ and it asymptotically vanishes when $B(0) \rightarrow \infty$. Equation (9)
with constant $B$ becomes ordinary linear differential equation. It can easily be solved by the standard methods
\cite{10} in the intervals of $r$ on which $G$ has a constant sign.  In the case $G >0$ the general solution
which obeys the conditions $G'(0)=0$ has the form
\begin{equation}
G(r) = \frac{1}{B^2(0)} + A \frac{\sin(B(0) r)}{r},
\end{equation}
where $A$ is a constant. The condition $G(R)=0$ gives
\begin{equation}
A = - \frac{R}{B^2(0)\:\sin(B(0) R)},
\end{equation}
while the condition $G'(R)=0$ yields the following relation
\begin{equation}
B(0) R = c_0,
\end{equation}
where $c_0$ is a solution of the equation
\begin{equation}
\tan c_0 = c_0.
\end{equation}
Equation (19) has infinitely many solutions, but only for
\begin{equation}
c_0 = 4.493...
\end{equation}
the function $G(r)$ has positive values in the whole interval $ 0 \leq r <R$ as required. Comparison with the
numerical solutions shows that such approximate analytic solution is quite accurate already for $B(0) = 4$ and
of course it becomes more accurate with increasing value of $B(0)$.

The  charge  for the approximate solution (16) is given by the formula
\begin{equation}
\underline{Q} = \frac{5}{6} \frac{R^6}{c_0^3}
\end{equation}
(the integral over $r$ is elementary). Formula (15) taken for $r =R$
implies that $ \omega = B(0) + \underline{Q}/R$, and with the help
of (18), (21)
\begin{equation}
\omega = \left(\frac{5}{6}\right)^{1/6} \left[\sqrt{c_0}\:
\underline{Q}^{\:-1/6} + \frac{1}{\sqrt{c_0}}\:
\underline{Q}^{\:5/6}\right].
\end{equation}
The total energy $\underline{E}$ is calculated from formula (13). The integral is split into $\int_0^R$ and
$\int_R^{\infty}$. In the latter one only $B' = \underline{Q}/r^2$ does not vanish. Elementary integrations
yield the following formula
\begin{equation}
\underline{E} =
\left(\frac{5}{6}\right)^{1/6}\left[\frac{12}{5}\sqrt{c_0}\:
\underline{Q}^{\:5/6} + \frac{1}{\sqrt{c_0}}\:
\underline{Q}^{\:11/6}\right].
\end{equation}

We see from formulas (18)-(23) that the radius $R$, the  charge
$\underline{Q}$, and the total energy $\underline{E}$ decrease with
increasing values of $B(0)$ -- the Q-ball becomes smaller and
smaller. Its charge density ($\sim B G^2$) exactly vanishes outside
the radius  $ R = (6 c_0^3/5)^{1/6}\: \underline{Q}^{1/6}$.
Generally, the charge density has a maximum at the center ($r=0$),
and it monotonically decreases towards zero, provided that
$\underline{Q}$ is small enough. For larger values of
$\underline{Q}$ the charge density is maximal at a finite radius $r
>0$.

The approximate formulas (16) - (18) imply that the initial data $(B(0), G(0))$ for the pertinent trajectories
of the fictitious particle lie on the curve
\[ G(0) B^2(0) = 1- \frac{1}{\cos c_0} = 5.603...,
\]
($\cos c_0 = -0.217...$) which is plotted as the continuous line in Fig. 2. The thick dots denote the initial
data determined numerically. It is clear that there is a very good agreement if $B(0) \geq 4$ what corresponds
to $\underline{Q} \leq 0.018 $.

The numerical investigations of the Q-balls for smaller values of $B(0)$ (that is $B(0) <4$) have shown, rather
unexpectedly, that the initial data $(B(0), G(0))$ move to smaller values of $G(0)$, see Fig. 2. The largest
value of $G(0)$ is obtained for $B(0) \approx 1.5 $. When $B(0)$ is still decreased the points $(B(0), G(0))$
move to the `southern slope' of the valley. Close to the end of the $G(0)>0$ part of the dotted line in Fig. 2
the values of $B(0)$ slightly increase. The end of the numerically determined Q-ball line has the coordinates
$B(0) = 1.317...,\; G(0) =0\;$ ($\underline{Q} = 9.753...$, $\underline{E} = 73.237... $, $\omega=4.546...   $).

The charge $\underline{Q}$ and the total energy $\underline{E}$ of the Q-balls first grow as we move along the
dotted line, but later they slightly decrease, see Section 5. The maximal values $\underline{Q} =
 9.779...$ and $\underline{E}=73.47...$ for the Q-balls are
obtained for $G(0) = 0.056...$, $B(0)= 1.290...$. On the other hand, the radius $R$ always grows -- its value at
the end of the Q-ball line is $ R = 4.856... $. The very interesting region of $B(0) \approx 1.3$ with small
values of $G(0)$ is discussed in detail in Section 5.

\begin{center}
\begin{figure}[tph!]
\hspace*{1cm}
\includegraphics[height=6.5cm, width=11cm]{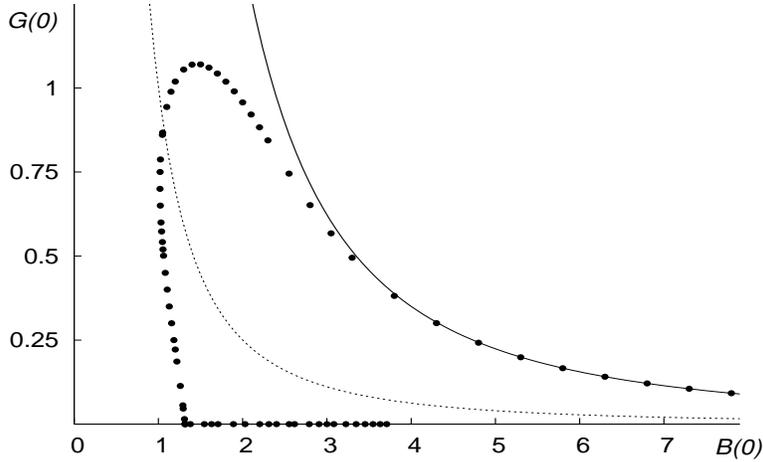}
\caption{The initial data for our numerical Q-ball and Q-shell solutions (represented by the thick dots). The
bottom of the valley is marked by the thin dotted line. The continuous curve is given by the equation $G(0)
B^2(0) = 5.603$ obtained from  the approximate analytic Q-ball solution (16). The thick dots on the line
$G(0)=0$ correspond to the Q-shells, and those with $G(0) >0$ to the Q-balls. The thick dots represent the
numerical solutions we have actually obtained - we do not suggest that the full set of initial data is
discrete.}
\end{figure}
\end{center}

The numerical solutions for the Q-balls show that for $B(0) \approx
1.3$ and small values of $G(0)$  the function $G(r)$ and the charge
density have maximal values at certain radius $r >0$, while at $r=0$
there is a local minimum, see Fig. 3. In particular, for $B(0) =
1.317...$, $G(0)=0$, that is at the end of the Q-ball line, the
charge density exactly vanishes at $r=0$. One may think that this
Q-ball is empty at the center. Such interpretation immediately
suggests the question whether there exist static Q-shells, for which
the charge density does not vanish  in a spherical shell with finite
internal and external radiuses. It turns out that the answer is in
the affirmative.

\begin{center}
\begin{figure}[tph!]
\hspace*{1cm}
\includegraphics[height=6cm, width=11cm]{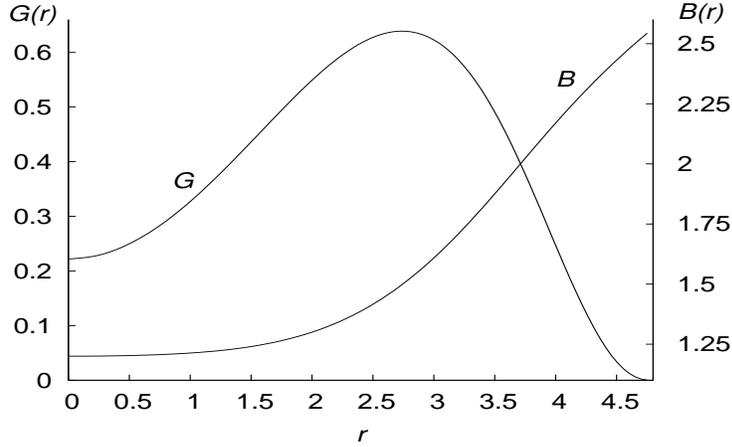}
\caption{Example of a large Q-ball solution. The function $G(r)$ has a local minimum at $r=0$. For this solution
$B(0) = 1.2, \;G(0)= 0.221...$, $R = 4.76...$, $B(R)=2.54...$, $\underline{Q} = 9.52...$, $\underline{E} =
71.2...$, $\omega= 4.54...$}
\end{figure}
\end{center}

\section{The Q-shells}

In the case of Q-shells the function $G(r)$ exactly vanishes when $r \leq r_0$ and also when $r \geq R >r_0$.
Hence, the electric charge is localized inside the spherical shell given by the condition $r_0 \leq r \leq R$.
The function $B(r)$ has constant value equal to $B(r_0)$ in the interval $0 \leq r \leq r_0$, and the Coulomb
form (15) when $r \geq R$. The corresponding trajectory of the particle from the mechanical analogy starts at
certain `time' $r =r_0>0$ from the point $(B(r_0),0)$ on the $B$-axis, tangentially to the axis, moves across
the valley and climbs the `northern slope' until $G'=0$ at a `time' $r_1 > r_0$ when it reaches a turning point.
Next, it again crosses the valley and comes back to the $B$-axis (tangentially to it) at the time $R > r_1$.
Solutions of Eqs. (8), (9) of that kind have been found numerically. An example is presented in Fig. 4.

The initial data for the trajectory of the fictitious particle in
the case of Q-shells have the coordinates $(B(0), 0)$, where
$B(0)=B(r_0)$.  They lie on the horizontal axis in Fig. 2.  The
mechanical analogy suggests that the Q-shell solutions exist for
arbitrarily large values of $B(r_0)$. The initial `time' $r_0$
(equal to the internal radius of the Q-shell) is determined by a
balance between the friction term $- 2 G'/r$ and the harmonic force
$- G B^2$ in Eq. (9). The point is that  the values of $B$ grow
monotonically with $r$ because of the `B-wind', and therefore the
harmonic force is larger when the fictitious particle returns to the
B-axis than during its flight to the `northern slope'. For this
reason the returning particle would just cross the B-axis and move
to the $G <0$ part of the $(B, G)$-plane were it not for the
friction which can take away the right amount of energy from the
particle (if $r_0$ is suitably adjusted), so that $G'=0$ exactly at
the `time' $R$ when the particle reaches the B-axis. Thus, the
friction term can not be neglected if the `B-wind' is present.

Note also that if we neglect the `B-wind' (as in the case of large
Q-shells discussed below) the friction would stop the particle
before it returns to the B-axis. Therefore, as far as the Q-shell
solutions are concerned the `B-wind' and the friction go together --
we may keep or neglect only both.

\begin{center}
\begin{figure}[tph!]
\hspace*{1cm}
\includegraphics[height=7cm, width=11cm]{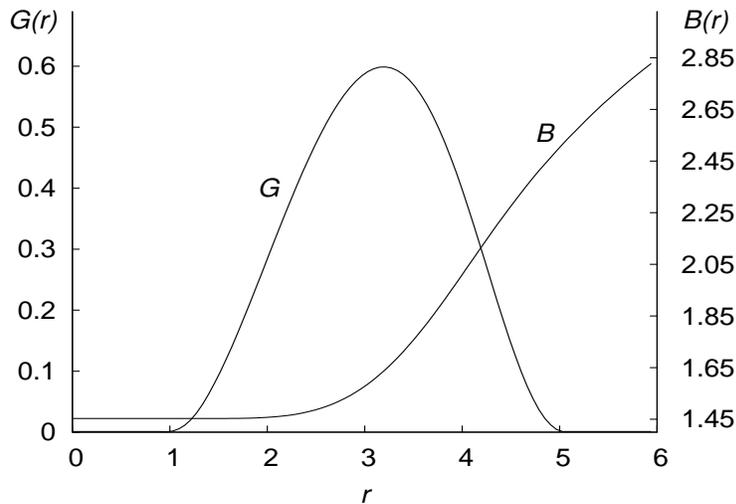}
\caption{Example of the Q-shell solution. The function $B(r)$ is constant for $r < r_0=1$, and almost constant
in the layer $1 \leq r \leq 2$ which lies inside the Q-shell. For this solution $ r_0 =1$, $B(0) =1.452...$, $R=
5.06...$, $B(R)=2.52...$,
 $\underline{Q} = 10.27...$,
$\underline{E} = 78.00...$, $\omega=4.55...$. }
\end{figure}
\end{center}

In the case of large Q-shells significant simplifications appear and it is possible to find approximate analytic
solutions. Because $r_0$ is large, the friction term in Eq. (9) may be neglected. Moreover, also $B(r_0)$ is
assumed  to be large.  In this region of the $(B, G)$-plane the valley is narrow and its northern slope is very
steep. Therefore, the fictitious particle will return to the B-axis rather quickly and the end point $(B(R),0)$
of its trajectory  will be close to the starting point $(B(r_0),0)$. Thus, we may assume that $B(r) \approx
B(r_0)$ in Eq. (9), i.e., that the `B-wind' is not important too. Numerical solutions show that indeed $B(r)$ is
almost constant inside large Q-shells, see Fig. 5.

\begin{center}
\begin{figure}[tph!]
\hspace*{1cm}
\includegraphics[height=7cm, width=11cm]{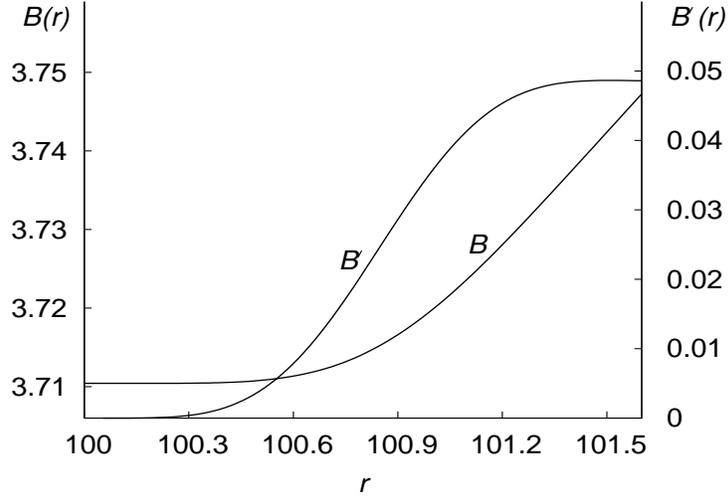}
\caption{Example of the function $B(r)$ with its derivative $B'(r)$ for a large Q-shell. For this solution
$r_0=100$, $B(r_0)=3.71...$, $R=101.69...$, $ B(R)=3.75...$, $B'(R)=0.04...$, $\underline{Q} = 501.8...$,
$\underline{E}= 7468.8...$, $\omega=8.68...$. At values of $r$ larger than shown in the picture the function
$B(r)$ becomes constant and $B'(r)$ decreases to zero. }
\end{figure}
\end{center}

The simplified form of Eq. (9) is as follows ($G>0$)
\begin{equation}
G''(r) = - B^2(r_0) G(r) +1.
\end{equation}
Both the `B-wind' and the friction have been neglected. Equation (24) has the following general solution
\begin{equation}
G(r) = \frac{1}{B^2(r_0)} + A \cos(B(r_0) r + \delta),
\end{equation}
where $A$ and $\delta$ are constants. The Q-shell solution has to obey the following conditions
\[
G(r_0) =0 = G'(r_0), \;\;\;\; G(R)=0=G'(R).
\]
In order to satisfy them we take \begin{equation} A = - \frac{1}{B^2(r_0)}, \;\;\;\; \delta = - B(r_0) r_0,
\end{equation}
and
\begin{equation}
B(r_0) (R- r_0) = 2 \pi.
\end{equation}
The relation (27) agrees very well with the numerical results already for $B(r_0) \geq 3.0$. For example, for
$B(r_0) = 3.083...$ we have  obtained  $B(r_0) (R- r_0) = 6.261...$.

Relation (27) gives $R$ provided we know  $r_0$ and $B(r_0)$.
$B(r_0)$ can be expressed by $r_0$ and $\underline{Q}$. To this end
we use the definition (14) in which $B(r) = B(r_0)$ and $G(r)$ is
given by formulas (25), (26). Elementary integration gives the
following formula
\begin{equation}
\underline{Q} = 3 \pi B^{-6}(r_0) ( x_0^2 + 2 \pi x_0 + 4 \pi^2/3 -
5/2),
\end{equation}
where
\[ x_0 =  r_0 B(r_0).\]

It remains to determine $r_0$. Equation (24) is not helpful here because it is invariant with respect to
translations of $r$. We will determine $r_0$ by minimizing the total energy $\underline{E}$ under the condition
that the  charge $\underline{Q}$ has a fixed value. Here it is convenient to use another formula for the energy,
namely
\begin{equation}
\underline{E} =  \underline{Q} B(R) + \frac{\underline{Q}^2}{R} + \int^R_{r_0}\! dr \: r^2 (G^{'2} + 2 |G|).
\end{equation}
It follows from the definition (13) by  applying  Eq. (8) and
integration by parts. Performing the (elementary) integration in
formula (29) and eliminating $R$ and $B(r_0)$  with the help of
formulas (27), (28) we obtain a rather lengthy formula for
$\underline{E}$ which contains only $\underline{Q}$ and $x_0$. In
the limit of large $x_0$ it can be written in the form
\begin{equation}
\underline{E} \cong (3\pi)^{1/6} \underline{Q}^{5/6} \left[ \frac{8}{3} x_0^{1/3} + ( \underline{Q} + \frac{8
\pi}{9} ) x_0^{-2/3} \right],
\end{equation}
where we have omitted terms of the order $x_0^{-5/3}$ or smaller. The minimum of $\underline{E}$ is obtained for
\begin{equation}
x_0 = \frac{3}{4} \underline{Q} + \frac{2\pi}{3}.
\end{equation}
Using this value of $x_0$ we find that in the limit of large $\underline{Q}$
\begin{equation}
\underline{E} \cong \sqrt{3}\: 2^{4/3} \pi^{1/6} \underline{Q}^{7/6},
\end{equation}
\begin{equation}
B(r_0) \cong \left(\frac{27 \pi}{16}\right)^{1/6}\underline{Q}^{1/6}.
\end{equation}
Then $r_0 \sim \underline{Q}^{5/6}$, and $R= r_0 + 2 \pi/ B(r_0)$. Comparison with numerical results shows that
formulas (30-33) are quite accurate already for $ \underline{Q} =  150$ -- in the case of $\underline{E},
B(r_0)$ they give numbers which differ from the corresponding numerical results by less than 1 percent, and for
$r_0$ it is less than 7 percent. Of course the discrepancies decrease when we go to larger values of $
\underline{Q}$.

\section{The transition from Q-balls to Q-shells }

The transition from Q-balls to Q-shells occurs at $\underline{Q}_* = 9.753...$. The largest Q-ball has the
radius $R_*= 4.85...$, and the initial data is $G(0) =0, \; B(0)= 1.317...$. For such small values of $B(0)$ the
`B-wind' can not be neglected. In consequence, the simple analytical approximations,  successfully applied in
the cases of small and large $\underline{Q}$, do not work. We have studied the pertinent solutions of Eqs. (8),
(9) numerically, concentrating mainly on the global characteristics: the energy $\underline{E}$ and the charge
$\underline{Q}$.

Let us first briefly describe the general structure of the solutions. On the Q-ball side, the minimum of the
function $G$  at $r=0$ gradually deepens until $G(0)=0$. The particle from the mechanical analogy starts its
motion from the `southern slope'. The `B-wind' carries it to the northern side of the valley (then $G(r)$
increases), where the particle makes a turn -- at that moment $G(r)$ has the maximal value. Next, it again
crosses the bottom of the valley and climbs the `southern slope' until it reaches the $G=0$ axis at the `time'
$r=R$, tangentially to the axis ($G'(R)=0$). For $r >R$ the particle moves along the B-axis. It slows down and
approaches the point $B(\infty)= \omega$ in the Coulomb way given by formula (15). On the Q-shell side, the
solutions have the shape presented in Fig. 4. Note the  influence of the `B-wind': $B(R)= 2.52...$ is
significantly larger than $B(0) = B(r_0) = 1.45...$.

Our main findings are presented in Fig. 6, where $\underline{E}$ and $\underline{Q}$ are plotted as the
functions of the radius $R$ (in the case of Q-shells it is the outer radius). It turns out that $R$ is a
convenient independent variable because it monotonically grows when we change the initial data along the thick
dotted line in Fig. 2 starting from the small Q-balls end ($G(0) >0, \; B(0) >>1$). Rather surprisingly, close
to $R_*$ the total energy and the  charge of the large Q-balls decrease when $R$ increases. The same is true for
the small Q-shells with the outer radius  slightly above $R_*$. Thus, the functions $\underline{E}(R), \;
\underline{Q}(R)$ have one local maximum and one local minimum. As far as we can see from our numerical data the
maxima occur at the same values of $R$. The same seems to be true for the minima. Note that the variations of
$\underline{E}(R), \; \underline{Q}(R)$ in this region of $R$ are rather small.

We know that $\underline{E}(R)\sim \underline{Q}^{5/6} $ for small Q-balls (i.e., when $ \underline{Q} <<
\underline{Q}_*$), and that $\underline{E}(R)\sim \underline{Q}^{7/6} $ for large Q-shells ($ \underline{Q} >>
\underline{Q}_*$). Using the numerical results summarized in Fig. 6 we have constructed the plot of
$\underline{E}$ versus $\underline{Q}$ for $\underline{Q}$ in the interval [9.73, 9.78]. It has the zigzag shape
schematically shown in Fig. 7. The two spikes correspond to the maxima and minima of $\underline{E}(R)$ and  $
\underline{Q}(R)$.  The presence of spikes in the plot of energy versus charge was found in other models with
Q-balls long time ago, see, e.g., \cite{11}.

\begin{center}
\begin{figure}[tph!]
\hspace*{1cm}
\includegraphics[height=8cm, width=13cm]{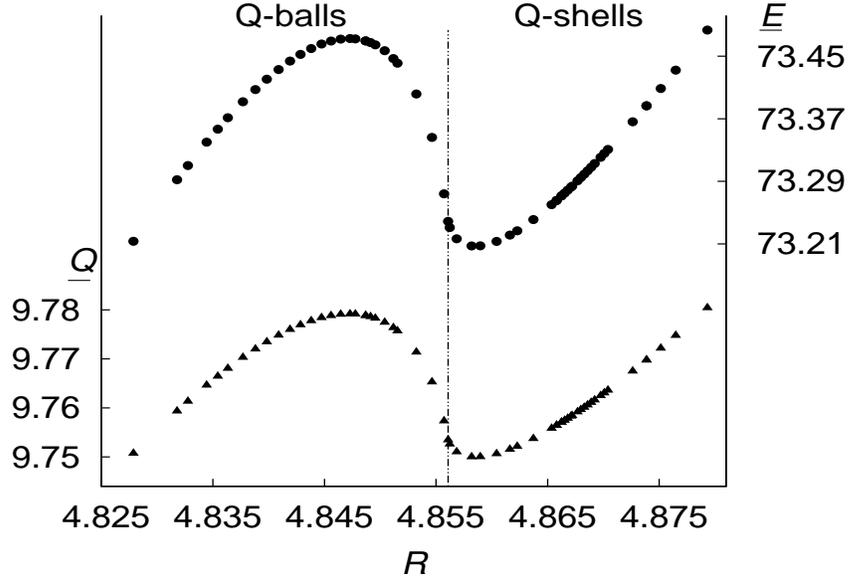}
\caption{The plot of the energy $\underline{E}$ (the thick dots) and of the total charge $\underline{Q}$ (the
triangles) versus the radius $R$ in the transition region. In the case of Q-shells $R$ is the outer radius.}
\end{figure}
\end{center}

We expect that the spikes are infinitely narrow. Our argument is
based on the formula
\[
\frac{d \underline{E}}{d \underline{Q}} = \frac{d \underline{E}/d
R}{d \underline{Q}/d R}.
\]
If the maxima of $\underline{E}(R), \; \underline{Q}(R)$ are located
at $R_1, \: R_2$, respectively, then close to them
\[
\frac{d \underline{E}}{d R} = \frac{1}{2}(R-R_1)^2
\left[\left.\frac{d^2 \underline{E}}{d R^2}\right|_{R_1} +
\frac{1}{3}\left.\frac{d^3 \underline{E}}{d R^3}\right|_{R_1}(R-R_1)
+...\right],
\]
\[
\frac{d \underline{Q}}{d R} = \frac{1}{2}(R-R_2)^2
\left[\left.\frac{d^2 \underline{Q}}{d R^2}\right|_{R_2} +
\frac{1}{3}\left.\frac{d^3 \underline{Q}}{d R^3}\right|_{R_2}(R-R_2)
+...\right],
\]
where $d^2 \underline{E}/d R^2|_{R_1}, \; d^2 \underline{Q}/d
R^2|_{R_2}$ are different from zero (and <0). If $R_1 \neq R_2$ the
derivative $d\underline{E}/ d \underline{Q}$ would become infinite
at the point $R=R_2$. The numerical results do not show any such
increase of $d\underline{E}/ d \underline{Q}$, hence  $R_1 = R_2$.
\begin{center}
\begin{figure}[tph!]
\hspace*{1cm}
\includegraphics[height=9cm, width=12cm]{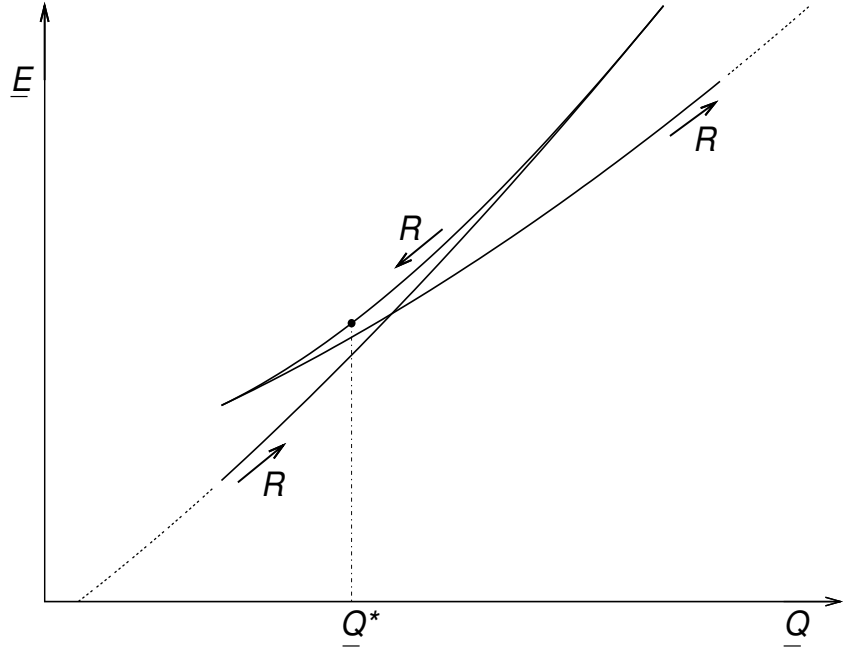}
\caption{The schematic picture of the zigzag in the plot
$\underline{E}(\underline{Q})$. The arrows point in the direction of
growing $R$. The dot on the uppermost line corresponds to the
solution which separates the Q-balls from the Q-shells ($B(0)=
1.317...$, $G(0)=0$). The upper spike corresponds to the maxima of
$\underline{E}$ and $\underline{Q}$, the lower one to the minima. }
\end{figure}
\end{center}
In consequence, $d\underline{E}/ d \underline{Q}$ is a continuous
function of $R$ at $R_1$. This means that the direction tangent to
the curve $\underline{E}(\underline{Q})$  changes infinitesimally
when we infinitesimally cross the maximum. Completely analogous
reasoning applies to the lower spike in Fig. 7 which corresponds to
the minima of $\underline{E}(R), \; \underline{Q}(R)$.

\section{Summary and remarks}

1. Let us summarize our results. We have shown with the help of the mechanical analogy and of the numerical
calculations that in the $U(1)$ gauged signum-Gordon model there exist compact, electrically charged non
topological solitons. They have the form of the Q-balls if the modulus of the total electric charge is smaller
than $Q_* = 4 \pi \underline{Q}_*/q$, where $\underline{Q}_* = 9.753...$ and $q>0$ is the electromagnetic
coupling constant. When $|Q|
> Q_*$ we have found the Q-shells which are completely empty
inside: the energy and the electric charge densities exactly vanish in a finite ball around the center of the
soliton. While electrically charged Q-balls have been found also in other models \cite{5, 6, 7, 8}, the static
Q-shells appear only in the gauged signum-Gordon model. Our feeling is that this is connected with the fact that
the scalar field of the signum-Gordon model reaches its vacuum value $\psi=0$ on a finite distance (in the
parabolic manner), not only in the direction of the infinity but also in the direction of the origin. Of course,
also the electrostatic repulsion is an important ingredient -- it prevents the Q-shell form a collapse.

Furthermore, we have found the approximate analytic solutions of the field equations which agree very well with
the numerical results in the limits of very small and very large absolute values of the electric charge. This is
especially important in the latter case  because the purely numerical approach is not sufficient in order to
convincigly  tackle the limit $|Q| \rightarrow \infty$.

2. It is interesting to see in detail the change of the exponent from $5/6$ to $7/6$ in the dependence of the
energy $\underline{E}$ on the  charge $\underline{Q}$ when passing from the asymptotic regions of very small to
very large charges. Note that the Figure 7 reminds a first order phase transition in, e.g., van der Waals gas,
\cite{12}. We have not found a good analytic approximation for solutions in the intermediate region.

3.  The formula given at the end of Section 2 shows that  the second variation of the energy $\delta^2
\underline{E} $ is positive. Therefore the Q-balls and the Q-shells are stable against small radial
perturbations. In general however the question of their stability is rather complex one if one considers also
nonsymmetric perturbations.  First, they can perhaps decay by emitting small Q-balls and/or a scalar and
electromagnetic radiation.  We have checked on our numerical solutions  that a split into two objects of the
charge $\underline{Q}/2$ is energetically forbidden only for small Q-balls ($\underline{Q}$ < 1.62...). This of
course does not prove that such a decay of large Q-balls in a finite time is dynamically possible. Furthermore,
the evolution of the perturbed solitons can be very nontrivial because of complicated electromagnetic
interactions of currents and charges. Note also that the usual criterion for the global stability consisting in
excluding the possibility of evaporation by  emission of quanta of the scalar field \cite{1} is hard to apply in
our case because the signum-Gordon model can not be linearized if the scalar field is close to the vacuum field.
Therefore, an estimate of the rest mass of the quanta is not available -- it is a rather difficult problem which
belongs to non perturbative quantum field theory.

4. We have investigated only the simplest (elementary) Q-balls and Q-shells. The mechanical interpretation of
Eqs. (8), (9) suggests that there also exist solutions such that the  charge density has several maxima. In this
case the classical particle travels along the valley for a longer time climbing the `southern slope' more than
once until it finally settles on the B-axis. Also numerical investigations have shown solutions of this type.
The relation $\underline{E}(\underline{Q})$ for such complex Q-balls and Q-shells is not known as yet. Another
open problem is their stability against a decay into the elementary Q-balls and Q-shells.

\section{Acknowledgement}
This work is supported in part by the MNII grant SPB  189/6.PRUE/2007/7.

\end{document}